# Light Sources Based on Relativistic Ion Beams

E.G.Bessonov

P.N.Lebedev Physical Institute, Moscow, Russia

Possible parameters of Light Sources (LS) based on Backward Rayleigh Scattering (BRS) of laser photons on relativistic ion beams in storage rings are discussed. It was shown that the parameters of the LS based on the RHIC storage ring (dedicated to High Energy Physics now) satisfy the requirements of the $2^d$ generation LS. Using the ordinary or enhanced radiative cooling of ion beams in storage rings by broad-band lasers will permit to decrease essentially their emittance, dimensions, angular and energy spreads and convert them to the next generation of LS.

## 1. Introduction

Progress in various fields of natural science, medicine, biology, chemistry, nuclear physics, technology is closely connected with the development and use of LS based on relativistic electron beams. They present the most brilliant sources of monochromatic IR to X-rays photon beams with high degree directionality, varied kinds of polarization and smoothly tuned in to photon energies in a wide spectral region. This is achieved by the production of low emittance electron beams in accelerators and storage rings of varied energy, by the use of a variety of undulators which can be optimized to the special demands of a certain experiment [1]-[5]. These sources make possible basic and applied research in different fields that are not possible with conventional equipment. They are undulator radiation (UR) sources including Backward (Inverse) Compton scattering (BCS) sources and future BRS sources using different kinds of the emitting ions instead of electrons. These sources are both spontaneous incoherent UR sources, stimulated UR sources or free electron lasers (fel) and spontaneous coherent UR sources (prebunched fel) [1]-[2]. The development of LS's goes on in all these directions. Three Generations of the LS are developed. Forth Generation is in the process. One of the main motivations to build new generation LS is to have brighter source of photons in a wide spectral region. Brightness and brilliance are determined by the equations

$$B_{rn} = \frac{1}{\hbar\omega} \frac{\partial^2 P}{\partial\omega\partial O}, \qquad B_{rl} = \frac{1}{\hbar\omega \cdot s_{eff}} \frac{\partial^2 P}{\partial\omega\partial O}, \qquad (1.1)$$

where $P$ is the power of the emitted beam, $\hbar$ is the Planck constant, $\omega$ is the angular light frequency, $O$ is the solid angle, $s_{eff}$ is the effective source area. It is convenient to consider that flux refers to the *number of photons/s/0.1percent BW*, brightness refers to *photons/s/unit solid angle in mrad/ 0.1percent BW* and brilliance refers to *photons/s/unit solid angle in mrad/0.1 percent BW/unit sours area in mm$^2$*. Forth generation LS will have average brilliance $>10^{22}$ and peak one $>10^{30}$ at photon energies $> 1$ KeV.

If the emitted light beam is focused onto the sample, then the smaller the particle beam dimensions and divergence, the smaller the sport size and divergence of the light beam at the sample. The brilliance was chosen to be an absolute criterion for generations of LS. But it can obscure essential distinctions between particular machines which determine if the machine is suited for a given application. Existing IR and optical FELs and future fully coherent monochromatic long duration X-ray FELs are not included in any generation of LS. A full characterization of LS involves specification of the intensity, brilliance, polarization, spectrum, coherence, and time structure of the emitted radiation. Old and new generations of LS can complement each other.

Possible parameters of Light Sources (LS) based on Backward Rayleigh scattering of laser photons on relativistic ion beams in storage rings are discussed below.

## 2. Light Sources based on Backward Rayleigh scattering of laser photons on relativistic ion beams

Below we consider a configuration in which a laser beam is colliding head-on with the ion beam in a straight section of an ion storage ring. Both the laser and ion beams are focused to a waist at the center of the interaction region. The central frequency of the incoming and scattered laser photons in this case will be written as $\omega_{l,0}$ and $\omega_{sc,0}$ respectively. Since we are considering the case near resonance, we have $\omega_l \simeq \omega_{l,0}$, $\omega_{sc} \simeq \omega_{sc,0}$. Let $\hbar\omega_{tr}$ be the transition energy in the ion's rest frame between two electronic or nuclear states 1 and 2, $\hbar\omega_l$ and $\hbar\omega_{sc}$ be the energies of the incoming and the scattered laser photons in the laboratory frame, respectively. These quantities are related by

$$\hbar\omega_l = \frac{\hbar\omega_{tr}}{\gamma(1-\beta\cos\psi)}, \qquad \hbar\omega_s = \frac{\hbar\omega_{tr}}{\gamma(1-\beta\cos\theta)}, \qquad (1.2)$$

where $\gamma = \varepsilon / Mc^2 = 1/\sqrt{1-\beta^2}$ is the ion relativistic factor, $\varepsilon$ the ion energy, $M$ its mass, $\beta = v/c$, $v$ the ion velocity, $c$ the speed of light, $\psi$ and $\theta$ the angles between the initial and final photon velocities and ion velocity respectively. When the ion beam has an angular spread $\Delta\psi$ around $\psi = 0$, and relative energy spread $\Delta\gamma$ around an average value $\gamma_b$, the full bandwidth required for the incoming laser to interact with all ions is, in view of Eq. (1.2),

$$\Delta\omega / \omega_{l,0} = (\Delta\psi)^2 / 4 + \Delta\gamma / \gamma_b, \qquad (1.3)$$

where $\omega_{l,0} \simeq \omega_{tr} / 2\gamma_b$. We assume that the incident laser beam has a uniform spectral intensity $dI/d\omega = I/\Delta\omega$ in the frequency interval $\Delta\omega$ centered on $\omega_{l,0}$, where $I$ is the total intensity.

The cross-section of the Rayleigh scattering of photons by ion in the ground state is

$$\sigma_{\omega l} = \frac{g_2 \pi^2 c^2}{g_1 \omega_{tr}^2} \Gamma'_{2,1} g(\omega_{tr}, \omega'_l), \qquad (1.4)$$

where $g_{1,2}$ are the statistical weights of the states 1 and 2, $\Gamma'_{2,1} = 2r_e \omega_{tr}^2 f_{1,2} g_1 / cg_2 \ll \omega_{tr}$ is the probability of the spontaneous photon emission of the excited ion or the natural line width in the ion's rest frame, $g(\omega_{tr}^2, \omega'_l) = \Gamma'_{2,1} / 2\pi[(\omega'_l - \omega_{tr}^2)^2 + \Gamma'^2_{2,1}/4]$ is the Lorentzian ($\int g(\omega)d\omega = 1$), $r_e = e^2/m_e c^2$ is the classical electron radius, $e$ and $m_e$ are its charge and mass, $\omega'_l$ is the frequency of the scattered monochromatic laser wave in the ion frame, and $f_{1,2}$ is the transition strength [6], [7]. The cross-section (1.4) has a maximum $\sigma_{max} = \sigma\,|_{\omega'_0 = \omega'_l} = g_2 \lambda_{tr}^2 / 2\pi g_1$, where $\lambda_{tr} = 2\pi c/\omega_{tr}$ is the resonance wavelength in the oscillator coordinate system. In the case of broadband laser beam of the width $\Delta\omega_l / \omega_l > (\Delta\psi_b)^2 / 4 + \Delta\gamma_b / \gamma_b$ the averaged cross-section

$$\bar{\sigma} = \pi f_{1,2} r_e \lambda_{tr} (\omega_l / \Delta\omega_l), \qquad (1.5)$$

where $\Delta\psi_b$ and $\Delta\gamma_b$ are the angular and energy spreads of the ion beam.

Note that electrons bounded by nuclei are very high quality oscillators. The amplitude of oscillations of an electron bounded by nuclei at resonance is much higher than for free electron in the same laser beam. That is why the cross-section of the Rayleigh scattering is much higher (~10-15 orders for monochromatic and by about a factor 7 – 8 orders for broadband lasers considered later) then Compton one.

In the ion reference frame the laser radiation scattered by one ion $i$ has the energy $\varepsilon'_{sc,i}$ and zero momentum. In the laboratory reference frame the scattered radiation have both the energy and momentum. In this case the energy of the scattered radiation $\varepsilon_{sc,i} = \gamma \varepsilon'_{sc,i}$. The density of the laser wave energy in the ion reference frame is increased $(1+\beta)^2 \gamma^2$ times and the length of the laser beam is shortened $(1+\beta)\gamma$ times [8]. That is why the energy of the scattered radiation $\varepsilon_{sc,i}$ is $(1+\beta)\gamma^2$ times higher the energy $\Delta\varepsilon_l = \varepsilon_{l,b}\bar{\sigma}/S_l(1+D)$ being knocked out from the laser wave. Here $\varepsilon_{l,b} = P_l \Delta t_l$ is the energy of the homogeneous laser bunch, $P_l$ is the power and $\Delta t_l$ is the duration of the laser bunch, $S_l$ is its cross section area, $D = I/I_{sat}$ is the saturation parameter, $I_{sat} = [g_1/4(g_1+g_2)](\hbar\omega_{tr}^4 / \pi^2 c^2 \gamma\bar{\gamma})(\Delta\omega/\omega)$ is the saturation intensity. Thereby, the energy of radiation scattered by one contra propagated ion in the laser bunch is $\varepsilon_{sc,i} = \varepsilon_{l,b}(1+\beta)\gamma^2 \bar{\sigma}/S_l(1+D)$. The average power of the beam scattered by one ion $\bar{P}_{\gamma,sc} = \varepsilon_{sc,i}/T_{rev}$ and the average power of the laser beam $\bar{P}_{l,b} = \varepsilon_{l,b}/T_{rev}$ have the same relation $\bar{P}_{\gamma,sc} = \bar{P}_{l,b}(1+\beta)\gamma^2 \bar{\sigma}/S_l(1+D)$, where $T_{rev}$ is the ion revolution period. If we suppose that the laser and ion beams have the bunch lengths $l_{lb}$ and $l_{ib}$ less than the Rayleigh length $Z_R = 4\pi\sigma_{lb}^2/\lambda_l$, the number of ions in the ion bunches is $N_{i,b}$, the number of bunches in the ring is $N_b$, the frequency of the revolution of the ion bunch in the storage ring $f_{rev} = 1/T_{rev}$ then the average power of the BRS radiation

$$\bar{P}_{\gamma,b} = N_{i,b} N_b \bar{P}_{\gamma,sc} = (1+\beta)\gamma^2 N_{i,b} N_b \bar{P}_{l,b} \bar{\sigma}/S_{eff}(1+D). \qquad (1.6)$$

where $S_{eff}$ for the homogeneous cylindrical laser and ion beams is their maximal cross section area. For the Gaussian beam density distribution $S_{eff} = S_l + S_i = 2\pi(\sigma_{lb}^2 + \sigma_{ib}^2)$, $\sigma_{lb}^2$ and $\sigma_{ib}^2$ are the rms dimensions of laser and ion beams. In the last case we did not take into account at averaging the dependence of the saturation parameter *D* on the transverse position in the laser bunch.

The occupation probability for the excited state 2 for the steady state regime $t >> \tau_{sp}$ is $n_2 = [g_2/(g_1+g_2)]D/(1+D)$, where $\tau_{sp} = 1/\Gamma_{2,l}$ is the spontaneous decay time, $\Gamma_{2,l} = \Gamma'_{2,l}/\gamma$. The quantity $n_2/\tau_{sp} = n_2 \cdot \Gamma_{2,l}$ can be interpreted as the number of scattered photons per ion per unit time.

The damping times for ion oscillations in the storage ring in radial, vertical and longitudinal directions are

$$\tau_x = \tau_y = \frac{\tau_\varepsilon}{(1+D)} = \frac{2\varepsilon}{\overline{P}_{\gamma,sc}}, \tag{1.7}$$

The rms relative energy spread of the ion beam at equilibrium is given by

$$\sigma_\delta = \sqrt{1.4(1+D)\hbar\omega_{tr}/Mc^2}. \tag{1.8}$$

In the present case, where the interaction takes place in a dispersion-free straight section, the equilibrium rms horizontal emittance (product of the beam dimensions & angular spread) is found to be

$$\epsilon_x = \frac{3\hbar\omega_{tr}\overline{\beta}_x}{20Mc^2\gamma_b^2}, \tag{1.9}$$

where $\overline{\beta}_x$ is the average horizontal beta function in the interaction region. The equilibrium vertical emittance is obtained by replacing $\overline{\beta}_x$ by $\overline{\beta}_y$ [9] [10].

The BRS radiation like UR and BCS radiation is the radiation of a moving oscillator. Its properties are determined by the type of the undulator (in our case we have laser beam undulator) [11] - [13]. That is why we can use the results of the UR theory for scattered radiation. E.g., the spectral distribution of the average power of the BRS radiation of the relativistic ion beam can be presented in the form

$$\frac{d\overline{P}_{\gamma,b}}{d\xi} = \overline{P}_{\gamma,b} f(\xi), \tag{1.10}$$

where $f(\xi) = 3\xi(1-2\xi+2\xi^2)$, $\xi = \omega_{sc}/\omega_{sc,\max}$, $\omega_{sc,\max} = \omega_{sc}|_{\theta=0} = (1+\beta)^2\gamma^2\omega_l$, $1/(1+\beta)^2\gamma^2 \leq \xi \leq 1$, $\int f(\xi)|_{\gamma>>1} d\xi = 1$. As opposed to UR, the frequency of the BRS radiation does not depend on the intensity of the laser wave and it is emitted on the first harmonic only.

The equation (1.10) is valid for the arbitrary polarised laser beam. The frequency band $\Delta\xi$, the corresponding angular $0 \leq \Delta\theta \leq \theta_\xi << 1$ and solid angular range $O_\xi = \pi\theta_\xi^2 = \pi\Delta\xi/\gamma^2$ of the radiation passing through the given collimator aperture, according to (1.2), are related by the angle $\theta_\xi \simeq \sqrt{\Delta\xi}/\gamma$ if the angular spread of the ion beam $\Delta\psi < \theta_\xi$. In this case the average brilliance of the BRS source (1.1) for Gaussian beams is

$$\overline{B}_{rl}|_{\theta<<1/\gamma_b} \simeq \frac{\Delta\xi}{\hbar\omega_{s,\max}s_{eff}O_\xi}\frac{d\overline{P}_{\gamma,b}}{d\xi}|_{\xi=1} = \frac{3 \cdot 10^{-6}\gamma^2\overline{P}_{\gamma,b}}{\pi s_{eff}\hbar\omega_{s,\max}} \tag{1.11}$$

in ph/s independent on BW/area/unit solid angle, where $s_{eff} = 2\pi\sigma_{eff}^2$, $\sigma_{eff} = \sigma_l\sigma_i/\sqrt{\sigma_l^2 + \sigma_i^2}$. Factor $10^{-6}$ in (1.11) takes into account the fact that the solid angle is measured in $mrad^2$.

Note that the brightness in this case does not depend on the value of the energy spread of the photon beam $\Delta\xi$. The angular spread of the ion beam, according to (1.2), must be $\Delta\psi < \theta_\xi$. The energy spread and the solid angle of the scattered beam in this case are determined by the diameter of the light collimator.

**Example.** Suppose that the Brookhaven RHIC [14] is used for which the length of the orbit $C = 3.833$ km, $\gamma = 97$, $N_{i,b} = 2 \cdot 10^9$, $N_b = 57$, the bunch spacing is $t_b = 224$ ns ($ct_b = 67.2$ m), the ion revolution frequency $f_{rev} = 78.32$ kHz, effective ion bunch length $\sigma_{i,b} = 20$ cm, emittance $\epsilon_x = 4 \cdot 10^{-7}$ mrad, beta function at the interaction of laser-ion beams $\overline{\beta}_x = 1$ m, rms ion beam dimension $\sigma_x = 6.3 \cdot 10^{-2}$ cm, the angular spread $\Delta\theta_{i,b} = \sqrt{\epsilon_x/\overline{\beta}} = 0.63 mrad$, relative energy spread $\Delta\gamma/\bar{\gamma} = 2.5\cdot10^{-3}$. Transition between the ground $(2s^22p^3)^4S3/2$ and excited $(2s^22p^4)^4P3/2$ states in the $N$-like xenon ions $^{54}_{129}\text{Xe}^{47+}$ is used. The transition parameters are: $g_1 = g_2 = 4$, $\hbar\omega_{tr} = 608.44$ eV, $\lambda_{tr} = 2.04\cdot10^{-7}$ cm, $f_{1,2} = 8.9\cdot10^{-2}$, $\bar{\sigma} = 6.26\cdot10^{-18}\text{cm}^2$ [8]. To excite the ions the laser beam with the wavelength $\lambda_l = 3954$ Å is used. In this case in the forward direction $\theta = 0$ the scattered beam hardness and intensity has maximum. The laser photon energy $\hbar\omega_l = 3.14\ eV$ maximum energy of scattered photons is $\hbar\omega_{s,\max} = 118$ keV and average one is $\hbar\bar{\omega}_s = 59$ keV, which is in the hard X-ray region. With the Rayleigh length $Z_R = 12.7$ m, the laser cross section is the same as the ion beam cross section at the beam waist location ($\sigma_{lb}^2 = \sigma_{ib}^2$). However, since $\in_x > \in_l$, the ion beam cross section is larger than the laser one away from the waist. To minimize this effect the length of the laser bunch $l_{lb} = 2$ m is used. We choose $D = 1$, laser intensity $I_l = I_{sat} = 2.87\cdot10^8$ W/cm$^2$, the duration of the laser bunch $\Delta t_l = l_{lb}/c = 6.67\cdot10^{-9} s$, the energy of the laser bunch $\varepsilon_{l,b} = P_l\cdot\Delta t_l = 2\pi\sigma_{lb}^2 I_{sat}\Delta t_l = 47.7$ mJ per pulse and the average laser beam power $\bar{P}_{l,b} = 213$ kW, which should be feasible with a free electron laser in an intracavity configuration. Gaussian distribution of the laser and ion density beams is used.

In this case $\Gamma'_{2,1} = 1.43\cdot10^{12}$, the laser photons flow is $\dot{n}_{l,ph} = I_{l,sat}/\hbar\omega_l = 5.7\cdot10^{26}$ ph/s/cm$^2$, the laser photon density $n_{l,ph} = \dot{n}_{l,ph}/c = 1.9\cdot10^{16}$, the saturation length $l_{sat} = 1/(1+\beta)n_{l,ph}\bar{\sigma} = 4.2$ cm and the spontaneous decay length $c\tau_{sp} \doteq \gamma\ 12,04$ cm are much less than the interaction length, the effective cross section area of the laser and ion beams interaction $S_{eff} = 4\pi\sigma_{ib}^2 = 4\pi\sigma_{lb}^2 = 4.99\ mm^2$, the effective light source area $s_{eff} = 1.25$ mm$^2$. The average power of the BRS radiation emitted by ion beam, according to (1.6), $\bar{P}_{\gamma,b}|_{\Delta\xi=1} = 1.788\cdot10^{23}\ eV/s = 28.7 kW$. The average brilliance of the LS, according to (1.11), $\bar{B}_{rl}|_{\Delta\xi=10^{-3}} = 1.09\cdot10^{16}\ ph/s\cdot mm^2\cdot10^{-3} BW\cdot mrad^2$, the pulse brilliance of the LS is $B_{rl}|_{\Delta\xi=10^{-3}} = \bar{B}_{rl}|_{\theta\leq\theta_\xi}\cdot q = 3.66\cdot10^{17}$ ph/s/mm$^2$/mrad$^2$/01%BW/, where $q = C/N_b\cdot l_{lb} = 33.6$ is the filling factor. The total average X-ray photon production rate is $\dot{\bar{N}}_{\gamma,b} = \bar{P}_{\gamma,b}f(\xi)\Delta\xi/\hbar\bar{\omega}_{sc,\max}|_{\Delta\xi=10^{-3},\xi=1} = 4.55\cdot10^{15}\ ph/s$. The corresponding pulse value is $q=33.6$ times bigger. The radiation emitted in the frequency band $\Delta\xi = 10^{-3}$ propagates in the angular band $\theta_\xi = 0.326\ mrad$ and in the solid angle $O_\xi = 0.33\ mrad^2$. They are about 2 and 4 times lesser than ion beam angular spreads. It means that the above brightness is a little overestimated. The transverse damping times are $\tau_x = \tau_y = 14$ min, while $\tau_\varepsilon$ is twice as long. They are more than 10 times shorter than what can be achieved with other methods, such as the stochastic cooling. The equilibrium beam parameters are $\sigma_\delta = 1.2\cdot10^{-4}$, which are about 20 times smaller than the initial value, and $\varepsilon_x = 8\cdot10^{-14}$ mrad is 7 orders smaller. It means that the beam dimensions, angular and energy spreads of the ion beam can be essentially decreased. Such a way the brightness of the LS based on BRS can be essentially increased.

## 3. Conclusion

1) The requirements to the angular spread $\Delta\theta_b < \sqrt{\Delta\xi}/\gamma$ of the ion beam in the BRS source is much less then to electron one emitting in magnetic undulators at the same hardness of the photon beams as the factor $\gamma$ for ions is much less then for electrons, the emitting wavepackets are much longer and the degree of monochromatisity can be much higher;

2) Opposed to UR, the frequency of the BRS radiation does not depend on the intensity of the laser wave. It is emitted on the first harmonic only. Background is lesser;

3) In the BRS source the total power of the synchrotron radiation of ions from bending magnets is much less than the BRS radiation. It means that the necessity in the powerful RF equipment in the BRS source fall away, the problem of excitation of the phase and betatron oscillations by quantum fluctuations of the radiation is disappear and superconducting bending magnets can be used;

4) There is no necessity in long straight sections for storage rings in the BRS source.

5) The kind of polarization and intensity of the BRS source can be changed quickly. The energy of the BRS LS can be continuously adjustable in a wide range by the change of the ion beam energy in the storage ring and the observation angle determined by the transverse position of the X-ray collimator.

6) The parameters of the considered LS based on the storage ring dedicated to high energy physics satisfy the requirements of the $2^d$ generation LS. Using the ordinary or enhanced radiative cooling of ion beams in storage rings by broad-band lasers will permit to decrease essentially the emittance of the ion beams, their dimensions, angular and energy spreads and convert them to the next generations of LS [1]-[5], [15].

7) Now the relativistic ion storage rings are used in the elementary particle physics. Later they can be used for production of BRS radiation. Dedicated BRS LS could be cheaper and more effective than considered here. Free Ion Lasers [16] and Quantum generators using both electronic and nuclear transitions based on relativistic cooled ion beams can be considered [1], [2].